\documentclass[twocolumn,showpacs,preprintnumbers,amsmath,amssymb]{revtex4}

\usepackage{graphicx}
\usepackage{dcolumn}
\usepackage{bm}
\usepackage{colordvi}
\usepackage{amsmath}
\usepackage{amssymb}
\usepackage{mathrsfs}

\begin{document}

\preprint{APS/123-QED}

\title{Dielectric relaxation of thin films of polyamide random copolymers}

\author{Natsumi Taniguchi}
\author{Koji Fukao}
\email{fukao.koji@gmail.com}
\altaffiliation[]{Corresponding author.}
\affiliation{%
Department of Physics, Ritsumeikan University, 
Noji-Higashi 1-1-1, Kusatsu, 525-8577 Japan
}%
\author{Paul Sotta}
\author{Didier R. Long}
\affiliation{%
Laboratoire Polym\'eres et Mat\'eriaux Avanc\`es
Unit\'e Mixte de Recherche CNRS/Solvay 5268,
Axel'One, 87, avenue des fr\`eres Perret 
F-69192 Saint Fons, France
}%

\date{\today}

\begin{abstract}
We investigate the relaxation behavior of thin films of a polyamide
 random copolymer, PA66/6I, with various film thicknesses using
 dielectric relaxation spectroscopy. Two dielectric signals are observed
 at high temperatures, the $\alpha$-process and the relaxation process
 due to electrode polarization (the EP-process). 
 The relaxation time of the EP-process has a 
 Vogel-Fulcher-Tammann type of temperature dependence, and the glass
 transition temperature, $T_{\rm g}$, evaluated from the EP-process agrees
 very well with the $T_{\rm g}$ determined from the thermal measurements. 
 The fragility index derived from the EP-process increases with
 decreasing film thickness.  
The relaxation time and the dielectric relaxation
 strength of the EP-process are described by a linear function of 
 the film thickness $d$ for large values of $d$, which can be regarded
 as experimental  
 evidence for the validity of attributing the observed signal to
 the EP-process. Furthermore, there is distinct  deviation from this
 linear law for thicknesses smaller than a critical value. 
This deviation observed in thinner films
 is associated with an increase in the mobility and/or diffusion
 constant of the charge carriers responsible for the EP-process.   
 The $\alpha$-process is located in a high frequency region than the
 EP-process at high temperatures, but merges with the EP-process 
at lower temperatures near the glass transition region.
 The thickness dependence of the relaxation time of the $\alpha$-process
 is different from that 
 of the EP-process. This suggests that there is decoupling
 between the segmental motion of the polymers and the translational
 motion of the charge carriers in confinement.  
    
\end{abstract}

\maketitle

\section{Introduction}
\label{introduction}

Amorphous materials exhibit glass transition behavior when cooled
from a high temperature to a low temperature under appropriate cooling
conditions~\cite{glass1}. At the glass transition temperature, $T_{\rm g}$, 
the motion of the $\alpha$-process is almost frozen, such that the
characteristic time of the motion extends to a macroscopic time scale. For
polymeric systems, the physical origin of the $\alpha$-process is
attributed to the segmental motion of the polymer chains. The freezing of the
$\alpha$-process can usually be explained by an anomalous increase in
the scale of the characteristic length of the dynamics when approaching
the glass transition 
temperature~\cite{Adam-Gibbs1965}.  
Dynamical heterogeneity is strongly correlated with this increase in the
scale of the characteristic length, and can be regarded as the most important
concept in elucidating the mechanism of the glass
transition~\cite{Berthier2010}. Glass transitions   
in confined geometry, such as thin polymer films and small molecules in
nanopores, have been widely investigated to determine the scale of the
characteristic length in the glass transition 
dynamics~\cite{MaKenna2005}. Recent 
measurements show that for thin polymer films there is a large deviation
in $T_{\rm g}$ and related dynamics from the bulk, 
albeit with some exceptions~\cite{Tress2010,Efremov2012}.

Being related to the heterogeneous dynamics near the glass transition
temperature, the correlation between two different modes of molecular
motion has been investigated~\cite{Chang1994,Ediger2000}. For example, 
decoupling between the translational and rotational motions is usually
observed in supercooled liquid states near $T_{\rm g}$. Similar
decoupling of the translational and rotational motions has also been
observed in confinement on the nanoscale~\cite{Napolitano2014}.

The dependence of the glass transition on the film thickness has been
investigated mainly for non-polar polymers, such as polystyrene, poly(methyl
methacrylate), and so on~\cite{Fukao2000,Napolitano2013}.  In contrast,
the glass transition dynamics of thin films of polar polymeric systems
have not yet been investigated intensively, although there have been
some reports on the dependence of glassy dynamics on the film thickness
for thin films of strong polar polymers such as
polysulfone~\cite{Labahn2009} and poly(bisphenol A carbonate)~\cite{Yin2012}.
Polyamides are one of these strong polar polymers~\cite{McCrum1967}. Because of
their industrial importance, many experimental investigations, including
dielectric measurements, have been carried out on polyamides over the
last fifty
years~\cite{Baker1942,Boyd1959,Laredo1997,Neagau2000,Lu2006,Laurati2012}. The
dielectric spectra of the polyamide family exhibit many processes, 
including $\gamma$-, $\beta$-, and 
$\alpha$-processes related to the molecular motion of polymer chains during
heating from a low temperature to a high temperature of up to 140$^{\circ}$C. 
These processes are generally attributed to local motions of the
methylene groups ($\gamma$-process), rotations of the amide
groups ($\beta$-process), or segmental motions
($\alpha$-process)~\cite{McCrum1967}. 
In addition to these three processes, other large dielectric signals are
observed in high-temperature and low-frequency regions, and should be
related to the motion of the charge carriers within the samples.
The DC conductivity process, the process related to the Maxwell-Wagner-Sillar
(MWS) interfacial polarization, and the electrode polarization process
(EP-process) are usually 
observed~\cite{BDS-book2003}. For polyamides, the charge 
carrier related to the three processes is mainly charged hydrogen, $i.e.$,
protons, which come from the amide linkage formed via hydrogen bonding
between the carbonyl group and the amino group~\cite{Baker1942}. 
For amorphous polymers, there is no interface between
the crystalline and amorphous phases. Hence, MWS polarization process is
not expected to be a major process in the low-frequency
and high-temperature region for amorphous polymers. In this case, the
EP-process might be 
more important than MWS polarization process. The molecular origin of the
electrode polarization is the partial blocking of charge carriers at the
sample-electrode interface. This leads to a separation of the positive and
negative charges, giving rise to additional polarization, as will
be further explained in Sec.~\ref{ep-process-theory}.

Recent dielectric measurements of Nylon 1010~\cite{Lu2006} show that the
dependence of the relaxation time of the EP-process on temperature can be
described by the Vogel-Fulcher-Tammann (VFT) law. This
suggests the possibility of monitoring the glassy dynamics
through investigation of the EP-process. Hence, thin films of polyamide
systems are expected to be suitable for the simultaneous investigation
of glassy dynamics and the motion of charge carriers, as well as the
correlation between the two processes. Because the translational motion
of charge carriers can be determined from the EP-process, polyamides can
be regarded as suitable systems for investigating the correlation
between the segmental motion of polyamide chains and the diffusion
motion of charge carriers in the polyamide matrix. 

In this study, we investigate the dielectric behavior of thin films of
an aromatic polyamide random copolymer with various film
thicknesses. Dielectric relaxation spectroscopy is used to elucidate the
dependence of the $\alpha$-process and the EP-process on the film
thickness, which could lead to a better understanding of the
relationship between the segmental motion and the motion of charge
carriers in confinement. Following this introduction, the experimental
details are given in Sec.~\ref{experiments}, and the theoretical
background, especially on the EP-process, is given in
Sec.~\ref{theory}. In Sec.~\ref{Res_Dis}, the experimental results from
dielectric relaxation spectroscopy are given, along with an analysis in
terms of the model developed by Coelho. Finally, a brief summary
is given in Sec.~\ref{conc_remarks}.

\section{experiments}\label{experiments}
 
\begin{figure}[b]
\includegraphics[width=8.5cm,angle=0]{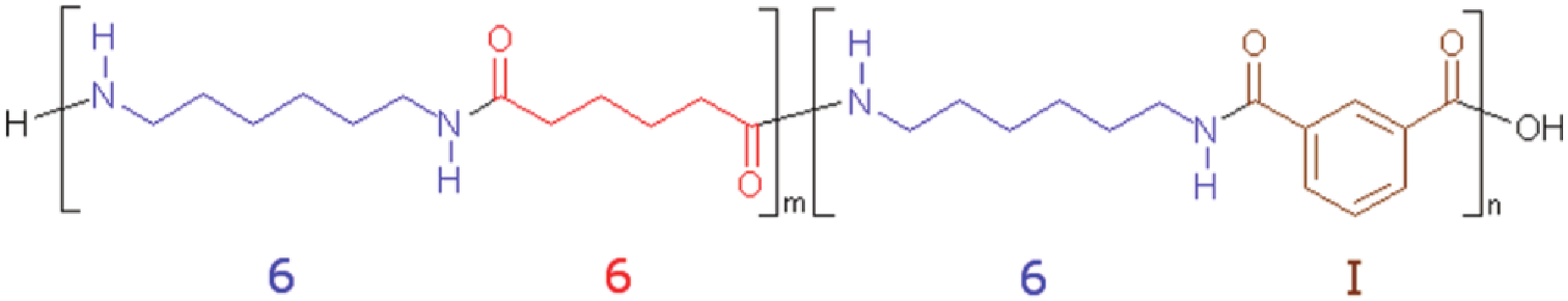}
\caption{%
The chemical formula of the amorphous polyamide copolymer PA66/6I.
}\label{pa666i-pic}
\end{figure}

\subsection{Samples}
The aromatic polyamide random copolymer PA66/6I was supplied by
Solvay~\cite{Laurati2012}. The weight-averaged molecular weight of the
copolymer is 
$M_{\rm w}$=1.5$\times$10$^4$, and the distribution of the molecular weight is
given by $M_{\rm w}/M_{\rm n}$=1.6. The chemical formula is
schematically shown in Fig.~\ref{pa666i-pic}. 
The glass transition temperature, $T_{\rm g}$, of the original material
was measured using differential scanning calorimetry (DSC), and was
found to be 357~K under heating at a rate of 10~K/min. 

The following procedures were carried out to prepare thin films for
dielectric measurements. The polyamide was first dissolved in
1,4-butanediol at 200$^{\circ}$C to obtain solutions of PA66/6I at
concentrations of 2 wt.\%, 1 wt.\%, and 0.5 wt.\%~\cite{Kiho1986}. The
solutions were then diluted at 160$^{\circ}$C with the same amount of
chloroform. Thin films were subsequently prepared on an aluminum
deposited glass substrate from the diluted solution using
spin-coating. Following that, aluminum was again vacuum-deposited to 
serve as an upper electrode. The sample was then annealed at 
160$^{\circ}$C for 12 hours before the dielectric measurements. 
The thickness of the film was controlled through the solution
concentration and the speed of rotation during spin coating. The
absolute film thickness was determined using atomic force microscopy 
measurements.

\subsection{X-ray scattering measurements}

X-ray scattering measurements were performed at SPring-8 BL40B2. A
simultaneous measurement system utilizing small-angle X-ray scattering (SAXS) 
and wide-angel X-ray scattering (WAXS) was used. The
measurement conditions were as follows: the camera lengths were 1757.4~mm and
90.0~mm for SAXS and WAXS, respectively; the X-ray wavelength was 0.9
$\AA$; the detector system was a CCD camera with an image intensifier for
SAXS, while that for WAXS was a flatpanel.

\subsection{Dielectric relaxation spectroscopy}
Dielectric relaxation spectroscopy (DRS) measurements were made using an
LCR meter (Agilent Technology, E4980A, 4292A) and Novocontrol AKB
analyzer. The measured frequency range was from 0.01~Hz to 2~MHz and the
temperature 
range was from 93~K to 423~K. The complex electric capacitance
$C^*_{\rm mes}(\omega)$ was obtained with the DRS measurements, where
$\omega=2\pi f$ 
and $f$ is the frequency of the applied electric field. Because of the
electric resistance $R$ of the electrode formed by the vacuum 
deposition of aluminum, there is an additional contribution to the
electric capacitance in the high-frequency 
region~\cite{Blum1995}. This contribution can be corrected using the
assumption that  
the sample condenser can be described as a serial circuit of the 
condenser $C^*(\omega)$ and $R$~\cite{Fukao2000}. 
Here, the following relation holds:
\begin{eqnarray}
C^*(\omega)&=&\frac{C^*_{\rm mes}+i\omega R|C^*_{\rm
 mes}|}{1-2\omega R\cdot{Im}(C^*_{\rm
 mes})+\omega^2R^2|C^*_{\rm mes}|^2}.
\end{eqnarray}
The complex electric capacitance $C^*(\omega)$ thus obtained is described
by $C^*\equiv C_0\varepsilon^*(\omega)$, where $C_0$ is the geometrical
capacitance and $\varepsilon^*(\omega)$ is the complex dielectric
permittivity of the sample. Here, $C_0=\varepsilon_0\frac{S}{d}$, where 
$\varepsilon_0$ is the dielectric permittivity {\it in vacuo}, $S$ is the area
of the sample or electrode, and $d$ is the thickness of the sample or the
distance between the two electrodes. In the present measurements,
$S$=8$\times$10$^{-6}$m$^2$. The complex dielectric permittivity is
given by:
\begin{eqnarray}
\varepsilon^*(\omega)=\varepsilon'(\omega)-i\varepsilon''(\omega), 
\end{eqnarray}
where $\varepsilon'$ and $\varepsilon''$ are the real and imaginary
parts of the complex dielectric permittivity, respectively.
The voltage applied to the samples for dielectric measurements was
selected as follows: 1.0~V for $d$=556~nm, 0.5~V for $d$=114~nm, 99~nm,
56~nm, 40~nm, and 0.2~V for $d$=20~nm. Hence, the applied electric field $E$
ranges from 2~MV/m to 12~MV/m for the present measurements.

\section{Theoretical models of the electrode polarization process} 
\label{theory}

\label{ep-process-theory}

\begin{figure}[b]
\includegraphics[width=8cm,angle=0]{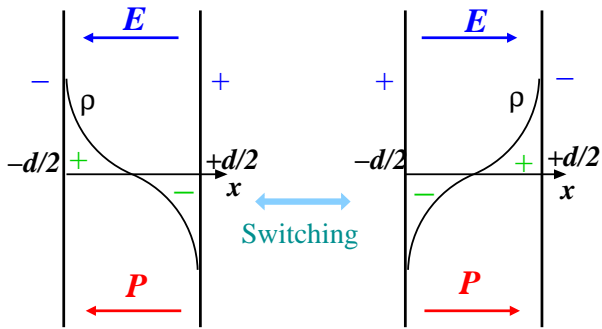}
\caption{%
Distribution of the charge between two electrodes under an applied
 electric field, with switching of the macroscopic dipole moment by
 reversal of the electric field. This figure has been modified from the
 original by Coelho. 
}\label{fig-ep-process}
\end{figure}

There are several theoretical models of the
EP-process~\cite{Macdonald1953,Coelho1983,Klein2006}.  
According to these theoretical models, the relaxation time of the electrode
polarization, $\tau_{\rm ep}$, and the relaxation strength of the 
electrode polarization, $\Delta\varepsilon_{\rm ep}$, can be described by a
linear function of the distance between the two electrodes, positive
and negative, which corresponds to the thickness of the thin film, $d$.

If there are mobile charge carriers in a sample, then the spatial
distribution of the  
charges changes depending on changes in the applied electric 
field. In the framework of the Debye theory, the dipolar relaxation can be
described as a process associated with viscous forces originating from
the matrix, where the charge carriers move randomly with thermal
fluctuations. In the absence of an applied electric field,  the charge
carriers are uniformly distributed in the sample, and the sample
is thus almost electrically neutral. In the presence of a DC electric
field, however, the charge carriers   
are separated depending on their polarity, and a heterogeneous charge
distribution appears within the sample. If the direction of
the applied electric field is reversed, then the charge carriers move in
such a way that a new 
heterogeneous charge distribution is reached, as shown in
Fig.~\ref{fig-ep-process}. For an AC electric field,  
the dynamics of the charge carriers have been modeled by
Coelho~\cite{Coelho1983}, and the complex dielectric permittivity has
been derived for applied electric fields with angular frequency
$\omega$ as follows:  
\begin{eqnarray}
\varepsilon^*(\omega)&=&\varepsilon_{\infty}\frac{1+i\omega\tau}{i\omega\tau+\dfrac{\tanh Y}{Y}},
\end{eqnarray}
where 
\begin{eqnarray}
Y&\equiv&\frac{d}{2L}\sqrt{1+i\omega\tau},\\
L&\equiv&\sqrt{D\tau}.
\end{eqnarray}
Here, $\tau$ is the relaxation time characterizing the relaxation
phenomena of the charge distribution or the electrode polarization upon
changes in the external applied electric field, $D$ is the diffusion 
constant of the charge carriers, and $L$ is the Debye length, $i.e.$,
the scale of the characteristic length of the electrostatic double layer near
the electrodes. In addition, the following relations among $\tau$, $\mu$,
$\sigma$, and $n_0$ are obtained:
\begin{eqnarray}
\tau^{-1}&=&\frac{\mu e
 n_0}{\varepsilon_{\infty}\varepsilon_0}=\frac{\sigma}{\varepsilon_{\infty}\varepsilon_0},\label{tau_1}\\
\sigma&\equiv&\mu en_0,\label{sigma_1}
\end{eqnarray} 
where $\sigma$ is the dc-conductivity, $n_0$ is the equilibrium
concentration of charge carriers, $e$ is 
the elementary electric charge, $\mu$ is the mobility of the charge
carrier, and $\varepsilon_{\infty}$ is the dielectric permittivity at very
high frequency.
Applying the Einstein relation leads to the following relation between
$D$ and $\mu$:
\begin{eqnarray}
D&=&\frac{\mu k_BT}{e},\label{diffusion_1}
\end{eqnarray}
where $k_B$ is the Boltzmann constant, and $T$ is the absolute
temperature~\cite{Einstein1905}. This equation is an expression of the
fluctuation-dissipation theorem~\cite{Kubo1991}. Hence, the Debye length
can be expressed as: 
\begin{eqnarray}
L&=&\frac{1}{e}\left(\frac{\varepsilon_{\infty}\varepsilon_0k_BT}{n_0}\right)^{1/2}. \label{Debye}
\end{eqnarray}

If $\omega\tau\approx$ 0 and $\delta\equiv d/2L\gg 1$, then the complex
dielectric permittivity can be described by: 
\begin{eqnarray}
\varepsilon^*(\omega)\approx\varepsilon_{\infty}\left(1+\frac{\delta}{1+i\omega\tau\delta}\right).
\end{eqnarray}
In real polymeric materials, there is a distribution of the relaxation
times because of the heterogeneity of the materials. The 
dielectric permittivity due to the EP-process can be
expressed as: 
\begin{eqnarray}
\varepsilon^*(\omega)&=&\varepsilon_{\infty}+\varepsilon_{\rm
 ep}^*(\omega),\\
\varepsilon_{\rm
 ep}^*(\omega)&=&\frac{\Delta\varepsilon_{\rm 
 ep}}{1+(i\omega\tau_{\rm ep})^{\alpha_{\rm ep}}},\label{eps_ep_1}
\end{eqnarray}
where 
\begin{eqnarray}
\Delta\varepsilon_{\rm ep}&=&\varepsilon_{\infty}\delta =
 \varepsilon_{\infty}\frac{d}{2L},\label{del_eps_ep}\\
\tau_{\rm ep}&=&\tau\delta=\frac{\varepsilon_{\infty}\varepsilon_0}{\mu en_0}\frac{d}{2L}.\label{tau_ep}
\end{eqnarray}
Using Eqns. (\ref{Debye}), (\ref{del_eps_ep}), and (\ref{tau_ep}), we can
obtain the following equations for 
the mobility $\mu(T)$ and the equilibrium concentration of charge
carriers $n_0(T)$:
\begin{eqnarray}\label{mu}
n_0(T)&=&\frac{4\varepsilon_0k_BT}{e^2\varepsilon_{\infty}d^2}(\Delta\varepsilon_{\rm 
 ep})^2,\\ 
\mu(T)&=&\frac{e\varepsilon_{\infty}d^2}{4k_BT}\frac{1}{\Delta\varepsilon_{\rm
 ep}\tau_{\rm ep}}, \label{n0}
\end{eqnarray} 
which relate $\mu(T)$ and $n_0(T)$ to the
experimentally observed values of $\Delta\varepsilon_{\rm ep}(T)$ and
$\tau_{\rm ep}(T)$. Here, $\tau_{\rm ep}$ is the relaxation time of the
EP-process.  
Furthermore, the diffusion constant of the charge carriers, $D(T)$, can be
determined using Eqs.~(\ref{diffusion_1}), (\ref{del_eps_ep}), and
(\ref{tau_ep}) from the values of $\Delta\varepsilon_{\rm ep}$ and 
$\tau_{\rm ep}$ as follows:
\begin{eqnarray}\label{diffusion}
D(T) &=&\frac{\varepsilon_{\infty}d^2}{4\Delta\varepsilon_{\rm ep}\tau_{\rm ep}}.
\end{eqnarray}

\begin{figure}
\includegraphics[width=8cm,angle=0]{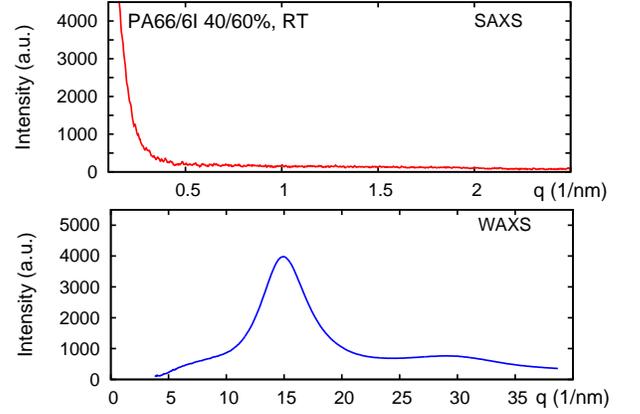}
\caption{%
X-ray scattering intensity with $q$ ranging from 0.1 nm$^{-1}$ to
 38~nm$^{-1}$ at room temperature for the amorphous polyamide copolymer 
 PA66/6I. The upper figure is for the SAXS region, and the lower one is for
 the WAXS region.}\label{x-ray}
\end{figure}

\begin{figure}
\hspace*{-0.3cm}
\includegraphics[width=8.8cm,angle=0]{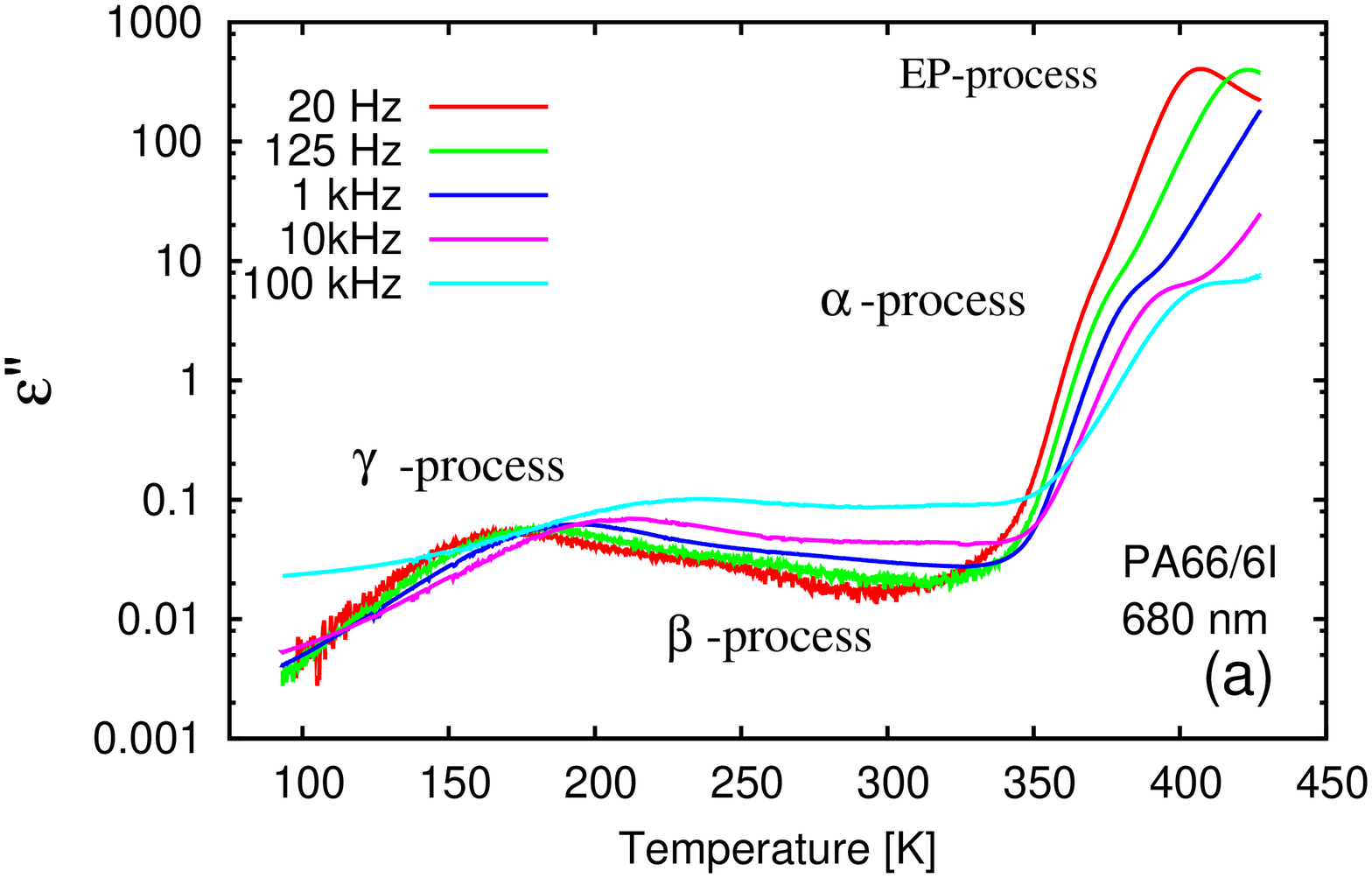}
\includegraphics[width=8.7cm,angle=0]{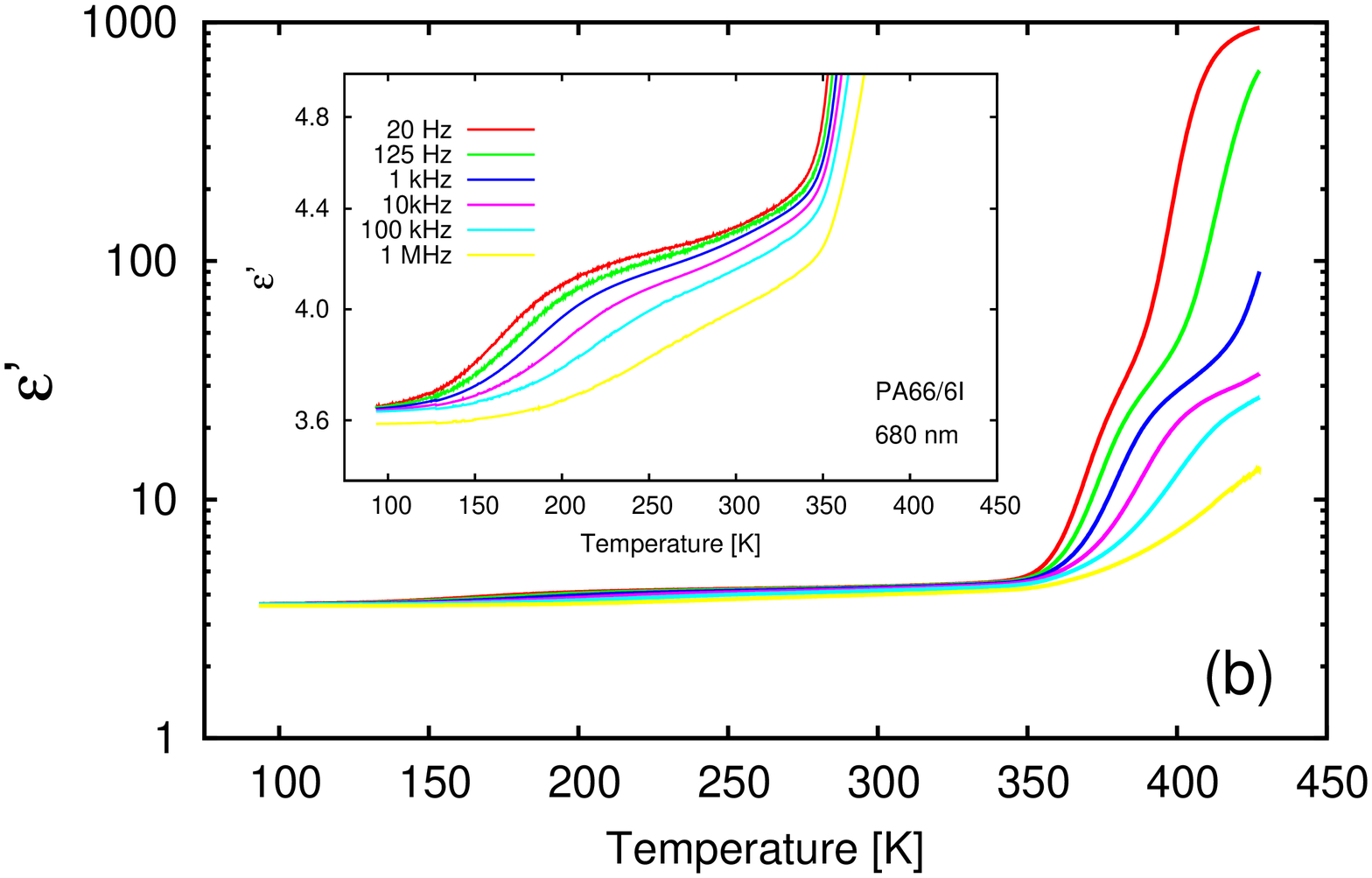}
\caption{%
The dependence of the complex dielectric permittivity on the temperature at
 various frequencies, for films of the amorphous polyamide copolymer
 PA66/6I with a thickness of 680~nm. (a) The dielectric loss
 $\varepsilon''$; (b) the real part $\varepsilon'$ and a 
a magnified image of $\varepsilon'$ (inset).}\label{temp-disp}
\end{figure}

\section{Results and Discussion}
\label{Res_Dis}

\subsection{X-ray scattering measurements}

Figure \ref{x-ray} shows the dependence of the X-ray scattering
intensity on the scattering vector $q$ at room temperature for the amorphous 
polyamide copolymer PA66/6I. The scattering profiles for both the SAXS and WAXS
regions are shown in Fig.~\ref{x-ray}. In the WAXS region, for $q>5$
nm$^{-1}$, there is a broad peak that is usually called an amorphous
halo. This peak originates from the short-range order of the structure. 
The amorphous halo is one typical scattering pattern for amorphous
materials. In the SAXS region, there is no peak, only a
continuous decay from $q=0$. Hence, we can infer that there is neither
a higher-ordered structure, such as a lamellar structure, nor a
crystalline structure in PA66/6I.

\subsection{Dielectric spectra in the temperature domain}

Figure \ref{temp-disp} shows the dependence of the complex
dielectric permittivity on the temperature, measured at various frequencies
for the amorphous polyamide
copolymer PA66/6I at a 680-nm thickness. In Fig.~\ref{temp-disp} (a),
there are several contributions to the dielectric permittivity depending on
the temperature and frequency. 
There is a dielectric loss peak due to the $\gamma$-process located at
around 170~K for 20~Hz, and there is also a broad contribution due to the
$\beta$-process between 220~K and 280~K. 
In the low-frequency and high-temperature region, there is a very large
contribution, which should be related to the motion of charge
carriers. Here, this behavior can be attributed to the EP-process and/or 
conductivity, as there should not be a crystal-amorphous 
interface in this amorphous polyamide copolymer. A detailed
discussion on attributing this to the EP-process is given in
Sec.~\ref{ep-process}. In addition to the three  
contributions, there is the $\alpha$-process as a shoulder of the
large peak of the EP-process.  In Fig.~\ref{temp-disp}(b),  
we can also see that there are signals in the real part of the dielectric
permittivity, which correspond to the four different dynamical contributions.
It should be noted that the observed data in Fig.~\ref{temp-disp} were 
obtained after heating above 100$^{\circ}$C. In other words, the results
in Fig.~\ref{temp-disp} should correspond to those for the dry
state~\cite{Laredo1997,Laurati2012}. 
In this paper, we will concentrate on the
$\alpha$-process and the EP-process.

\begin{figure}[b]
\hspace*{-0.3cm}
\includegraphics[width=9cm,angle=0]{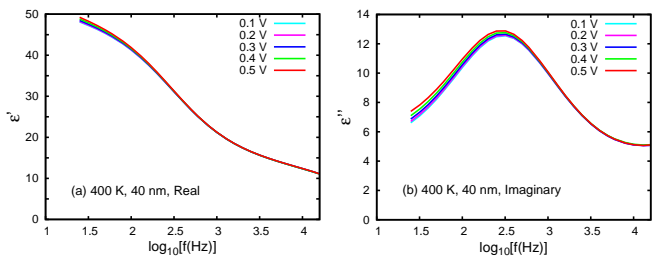}
\caption{%
The dependence of the real and imaginary parts of the complex dielectric
 permittivity on the frequency at 400~K for various applied voltages
 from 0.1~V to 0.5~V in thin films of the amorphous polyamide copolymer
 PA66/6I with a 40-nm thickness.
}\label{voltage}
\end{figure}

\begin{figure}
\includegraphics[width=8.3cm,angle=0]{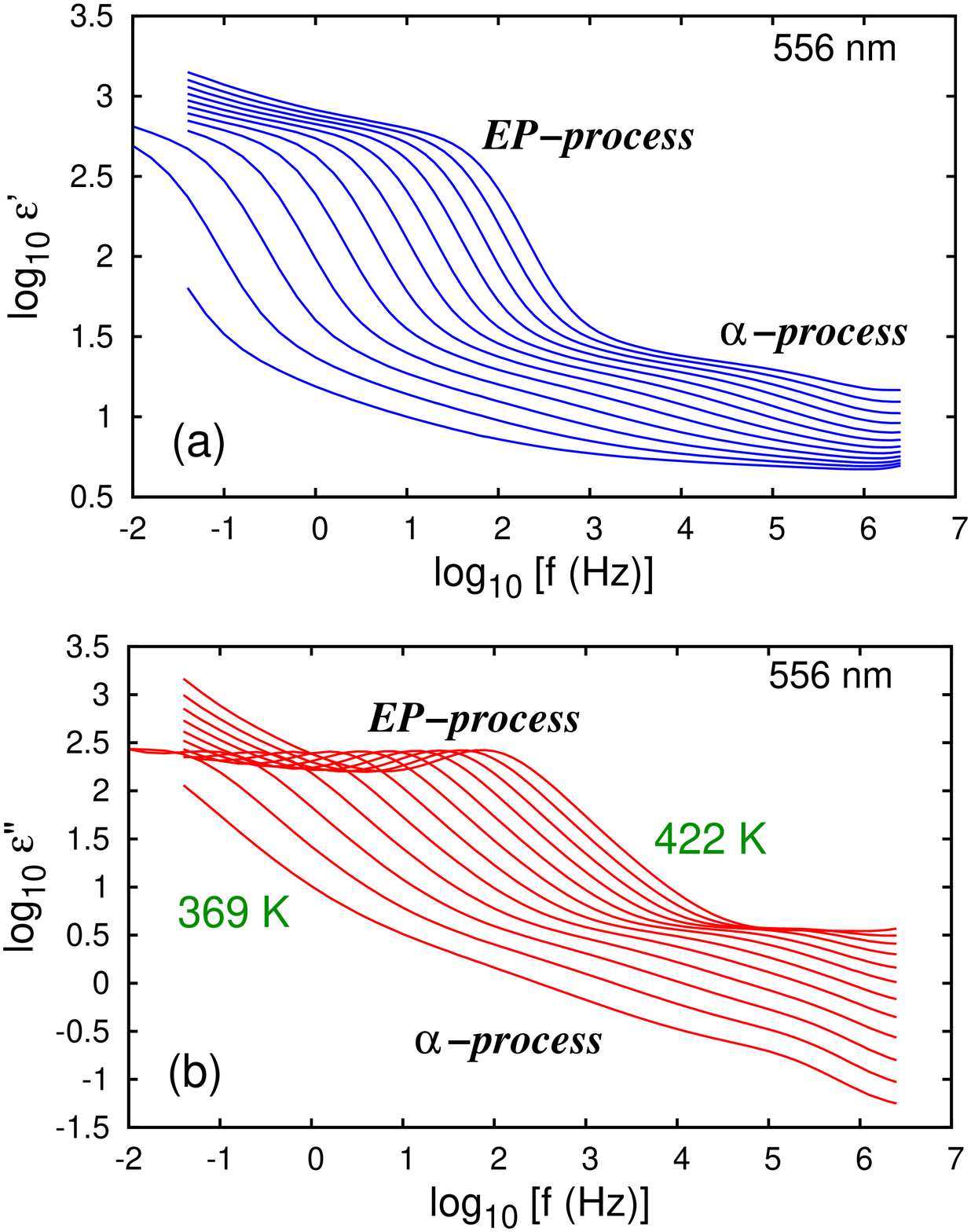}
\caption{%
The dependence of the complex dielectric permittivity on the frequency  
for thin films of the amorphous polyamide copolymer PA66/6I with a 
 thickness of 556~nm. (a) The real part of $\varepsilon^*$ and
 (b) the imaginary par of $\varepsilon^*$. The temperature
 ranges from 422.8~K to 369.5 K.} \label{eps-555nm}
\end{figure}

\subsection{Dielectric spectra in the frequency domain}

For strong polar materials, the EP-process may show a strong nonlinear
effect against the applied electric field. Figure \ref{voltage} shows
the real and imaginary parts of dielectric permittivity in the
frequnecy domain for various applied voltages from 0.1~V to 0.5~V in
thin films of PA66/6I with a 40-nm thickness. In Fig.~\ref{voltage},
the value of $E$ ranges from 2.5~MV/m to 12.5~MV/m, in which the
selected range of $E$ in the present measurements is almost
included. The curves observed for various applied voltages
are overlapped with each other, as shown in Fig.~\ref{voltage}. 
From these measurements, we can estimate possible maximum deviation due
to the change in electric field as follows: for the peak frequency of
the EP-process $f_{\rm ep}$ at a given temperature, possible deviation
$\Delta f_{\rm ep}/f_{\rm ep}$ is less than 
0.02, and for the peak value of $\varepsilon''$ of the EP-process,
deivation is less than 3 \%. Therefore, the observed change in peak
position and height of the EP-process with change in the film thickness is
much larger than these possible deviations. Therefore, it can be justified
that experimental results in this paper do not depend on the value of
$E$ for the range of $E$ selected in the present measurements. 

\begin{figure}
\includegraphics[width=5.5cm,angle=-90]{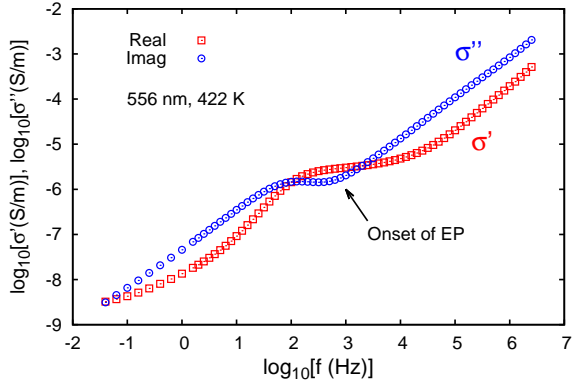}
\caption{%
The dependence of the logarithm of the real and imaginary part of
 complex conductivity observed at 422~K for thin films of PA66/6I with
 a 556-nm thickness. The arrow shows the onset of the EP-process.
}\label{cond_presen}
\end{figure}

\begin{figure}
\includegraphics[width=9cm,angle=0]{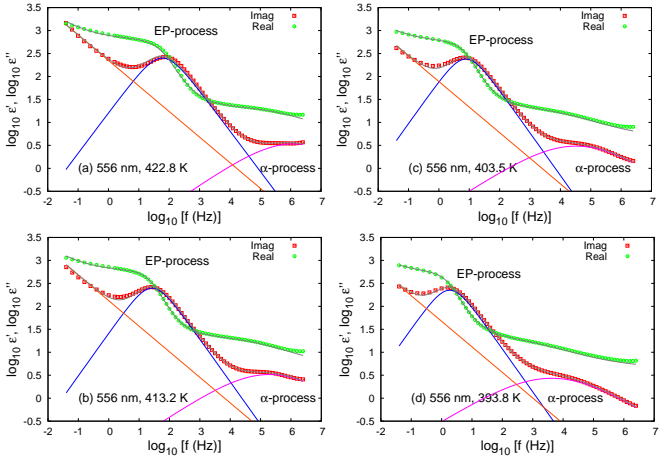}
\caption{%
The dependence of the real and imaginary parts of the complex dielectric
 permittivity on the frequency at various temperatures, (a)422.8~K,
 (b)413.2~K,  (c)403.5~K, and (d)393.8~K, for thin films of the
 amorphous polyamide copolymer PA66/6I with a 556-nm thickness showing
 ({\it red squares}) the  imaginary part $\varepsilon''$ and ({\it green
 circles}) the real part $\varepsilon'$. The three different
 contributions to the imaginary part are also shown: the orange line
 corresponds to the low-frequency component $\varepsilon_{\sigma}^*$, the
 blue line corresponds to the 
 EP-process, and the pink line corresponds to the $\alpha$-process. The
 curves for the three components and the two observed values of
 $\varepsilon'$ and $\varepsilon''$ were evaluated by fitting the
 observed values to Eq.~(\ref{com_eps}).}\label{eps-555nm-fit} 
\end{figure}

Figure \ref{eps-555nm} shows the dependence of the complex
dielectric permittivity on the frequency for thin films of the amorphous
polyamide copolymer PA66/6I with a 556-nm thickness. In
Fig.~\ref{eps-555nm}, different processes including the $\alpha$-process
and the EP-process contribute to the dielectric permittivity.
These components are readily observed over the temperature range from
369~K to 422~K. Furthermore, we also show the dependence of the complex
conductivity $\sigma^*(=\sigma'+i\sigma'')$ on the frequency for the
thin films. Here, $\sigma'$ and
$\sigma''$ are the real and imaginary parts of the complex conductivity,
respectively, and the values of $\sigma'$ and $\sigma''$ are evaluated
from the following relataions:
$\sigma'=\varepsilon_0\varepsilon''\omega$ and
$\sigma''=\varepsilon_0\varepsilon'\omega$.  Figure \ref{cond_presen}
shows the dependence of the logarithm of the real and imaginary parts of
the complex conductivity observed at 422~K for thin films of PA66/6I
with a 556-nm thickness. 
The value of $\sigma'$ approaches a constant value (plateau value)
at ca. 10$^4$Hz with decreasing frequency, although there is no exact 
plateau but with a small slope because of the overlap of the
$\alpha$-process. On the other hand, the value of $\sigma''$ shows the
signal of the onset of EP-process at 10$^3$~Hz. According to the usual
interpretation on the EP-process in the conductivity representation
proposed by Kremer and
coworkers~\cite{Serghei2009,Sangoro2012,Sangoro2014}, it is reasonable
to regard the 
``plateau'' of $\sigma'$ at 10$^4$Hz as the onset of dc-conductivity,
which should be required for the appearance of the EP-process.
If this interpretation is valid, the physical origin of the increase in
$\varepsilon'$ and $\varepsilon''$ in the low frequency region should
not be due to usual conductivity. 
Wang {\it et al.} reported that there is an increase in $\varepsilon'$ and
$\varepsilon''$ in the lower frequency region than the location of the
EP-process~\cite{Wang2013}.  Furthermore, they attributed this
contribution to imperfectness of the blocking at electrodes of the 
charge carriers which are responsible for the EP-process. This effect
causes a deviation from the expected values from the theoretical model
of the EP-process. Here, we adopt this interpretation for the increase
in $\varepsilon'$ and $\varepsilon''$ with decreasing frequency in
low-frequnecy region. In this case, the contributions in the
low-frequency region in $\varepsilon'$ and $\varepsilon''$ are not
directly related to the usual dc-conductivity. 

%

The frequencies at which the dielectric loss peaks due to the EP- and
$\alpha$-processes are located are shifted from the lower frequency region to
the higher frequency region with increasing temperature, as shown in
Fig~\ref{eps-555nm}.  
In order to investigate the dielectric properties of the components
separately, 
the observed dependence of $\varepsilon'$ 
and $\varepsilon''$ on the frequency was reproduced by the following
model function:
\begin{eqnarray}
\varepsilon^*(\omega)&=&\varepsilon_{\infty}+\varepsilon^*_{\sigma}(\omega)+\varepsilon^*_{\alpha}(\omega)+\varepsilon^*_{\rm
 ep}(\omega),\label{com_eps}
\end{eqnarray}
where $\varepsilon^*_{\rm ep}(\omega)$ is the complex dielectric
permittivity due to the EP-process described by Eq.~(\ref{eps_ep_1}),
$\varepsilon^*_{\alpha}(\omega)$ is the dielectric
permittivity due to the $\alpha$-process, and 
$\varepsilon^*_{\sigma}(\omega)$ is the dielectric permittivity due to the
imperfectness of the blocking at electrodes of the charge carriers.
For the component of $\varepsilon^*_{\sigma}$, 
we use simple power-law functions for both real and imaginary parts of
$\varepsilon^*_{\sigma}$: 
\begin{eqnarray}
Re[\varepsilon^*_{\sigma}](\omega)&=&
 A\omega^{-\tilde{m}}, \quad Im[\varepsilon^*_{\sigma}](\omega) =
 B\omega^{-\tilde{m}'}, 
\end{eqnarray}
where $A$, $B$, $\tilde{m}$, and $\tilde{m}'$ are constants. These functions work well,
as shown in Fig.~\ref{eps-555nm-fit}.
As for the $\alpha$-process, we adopt the 
Havriliak-Negami equation~\cite{HN1967}:  
\begin{eqnarray}
\varepsilon^*_{\alpha}(\omega)&=&\frac{\Delta\varepsilon_{\alpha}}{(1+(i\omega\tau_{\alpha})^{\alpha_\alpha})^{\beta_{\alpha}}},
\end{eqnarray}
where $\Delta\varepsilon_{\alpha}$ is the dielectric relaxation strength,
$\alpha_{\alpha}$ and $\beta_{\alpha}$ are shape parameters, and
$\tau_{\alpha}$ is the relaxation time of the $\alpha$-process.
In the present analysis, the parameter $\beta_{\alpha}$ is fixed at 
unity because of a reduction in the free fitting parameters.

Figure \ref{eps-555nm-fit} shows the dependence of $\varepsilon'$ and
$\varepsilon''$ on the frequency at four different temperatures, 422.8~K, 
413.2~K, 403.5~K, and 393.8~K, for a 556-nm-thick thin film of the amorphous
polyamide copolymer PA66/6I.  Calculated curves for the 
different components are also shown, and were obtained by fitting the
observed data to the 
model function, Eq.~(\ref{com_eps}). From Fig.~\ref{eps-555nm-fit}, we can
see that the 
observed dielectric permittivity is very well reproduced by the
present model function and, as a result, several important physical
quantities can be evaluated as a function of the temperature and film
thickness. 

\begin{figure}
\includegraphics[width=8.3cm,angle=0]{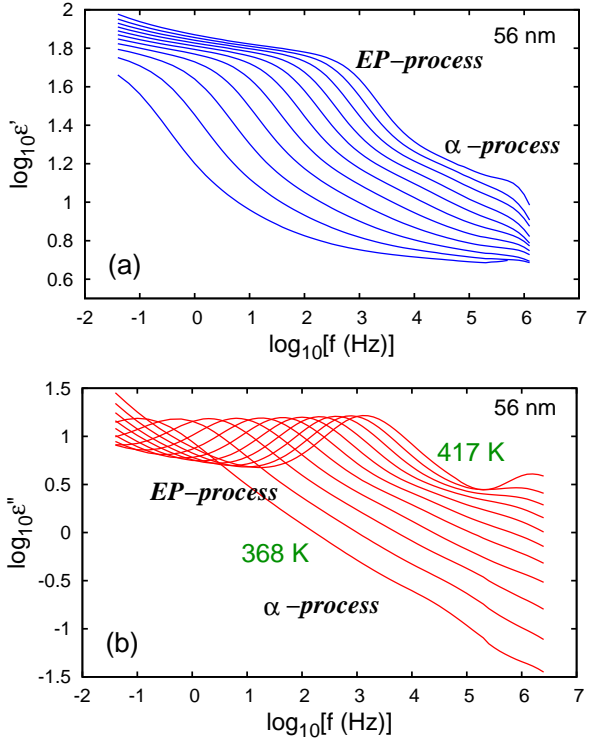}
\caption{%
The dependence of the complex dielectric permittivity on the frequency  
for thin films of the amorphous polyamide copolymer PA66/6I with a 
 thickness of 56~nm. (a) The real part of $\varepsilon^*$ and
 (b) the imaginary part of $\varepsilon^*$. The temperature
 ranges from 417.6~K to 368.7 K.
 }\label{eps-56nm}
\end{figure}

\begin{figure}
\includegraphics[width=8cm,angle=0]{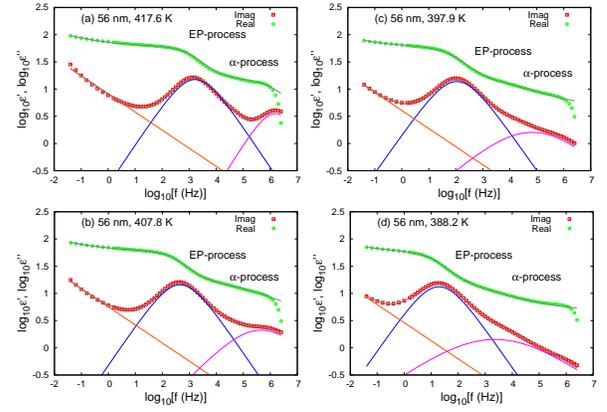}
\caption{%
The dependence of the real and imaginary parts of the complex
 dielectric permittivity on the frequency at various temperatures,
 (a)417.6~K, (b)407.8~K, (c)397.9~K, and (d)388.2~K, for thin films of
 the amorphous polyamide copolymer PA66/6I with a 56-nm thickness. The
 meanings of the symbols and curves are identical to those in
 Fig.~\ref{eps-555nm-fit}.  
}\label{eps-56nm-fit}
\end{figure}

Figure \ref{eps-56nm} shows the dependence of the complex
dielectric permittivity on the frequency for a 56-nm-thick thin film of
the amorphous polyamide 
copolymer PA66/6I. In Fig.~\ref{eps-56nm}, we can see that the peak
height of the EP-process in the curve of $\varepsilon''$~vs.~$f$ is much
smaller for the 56-nm-thick film than for the 556-nm-thick
film. Furthermore, the peak positions of the EP-process in the frequency
domain at a given temperature are clearly shifted to
the higher frequency region for the 56-nm-thick thin film, as compared
to the 556-nm-thick film. In addition, for the 56-nm-thick film, the
present model function again accurately reproduces the observed dielectric 
permittivity as a function of frequency for various temperatures, as shown
in Fig.~\ref{eps-56nm-fit}.

\subsection{The electrode polarization process}
\label{ep-process}

\begin{figure}
\includegraphics[width=8cm,angle=-90]{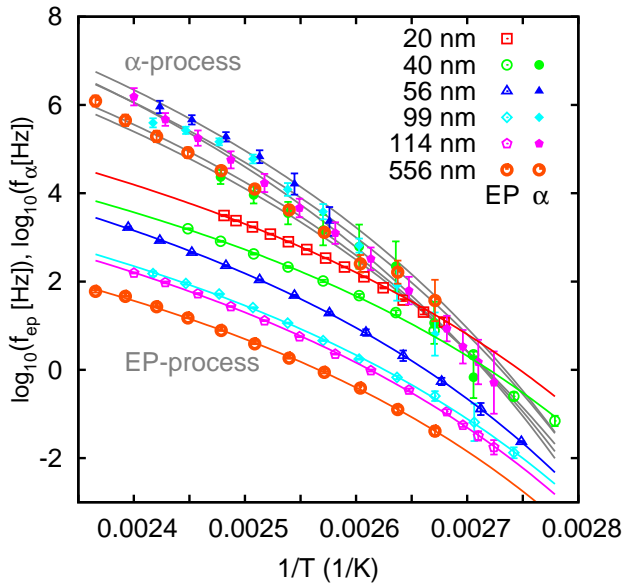}
\caption{%
Dispersion map of the relaxation rate $f_{\rm ep}$ of the EP-process and
 the relaxation rate $f_{\alpha}$ of the 
 $\alpha$-process, for thin films of the amorphous polyamide
 copolymer PA66/6I with various thicknesses ranging from 20~nm to
 556~nm. The values of 
$f_{\rm ep}$ and $f_{\alpha}$ are evaluated as $f_{\rm ep}=(2\pi\tau_{\rm
 ep}$)$^{-1}$ and $f_{\alpha}=(2\pi\tau_{\alpha}$)$^{-1}$,
 where
$\tau_{\rm ep}$ and $\tau_{\alpha}$ are the fitting parameters obtained
 by fitting the observed $\varepsilon^*$ to Eq.~(\ref{com_eps}). The
 curves are governed by the VFT law. The curves of the $\alpha$-process
 were obtained under the condition that the Vogel temperature $T_0$ was
 fixed at the value obtained by fitting the temperature
 dependence of $\tau_{\rm ep}^{-1}$ to Eq.~(\ref{vft_ep}). 
}\label{fmax-T-EP-1}
\end{figure}

Figure \ref{fmax-T-EP-1} shows the dependence of the
relaxation rate of the EP-process, $f_{\rm ep}$, on the temperature, for
thin films of the amorphous polyamide copolymer PA66/6I of various
thicknesses, ranging from 20~nm 
to 556~nm. The value of $f_{\rm ep}$ is evaluated from the relation 
$f_{\rm ep}=(2\pi\tau_{\rm ep})^{-1}$ with the best-fitted values of
$\tau_{\rm ep}$ obtained using Eq.~(\ref{eps_ep_1}). Here, the parameter
$\beta_{\rm ep}$ is fixed to 1, such that the resulting frequency $f_{\rm ep}$
is equal to the frequency at which $\varepsilon''$
exhibits a peak. In Fig.~\ref{fmax-T-EP-1}, we can see that
the relaxation rate of the EP-process has a stronger temperature
dependence than the Arrhenius type of temperature dependence.
As shown by the curves in Fig.~\ref{fmax-T-EP-1}, the temperature
dependence of $f_{\rm ep}$ is well reproduced by the VFT
law~\cite{Vogel1921,Fulcher1925a,Fulcher1925b,Tammann1926}:  
\begin{equation}\label{vft_ep}
\tau_{\rm ep}(T)=\tau_{\rm ep, 0}\exp\left(\frac{U}{T-T_0}\right),
\end{equation}
where 
$U$ is a positive constant, and $T_0$ is the Vogel temperature. 
At a given temperature, the relaxation rate of the EP-process increases
with decreasing film thickness. In other words, the relaxation time
decreases. The 
VFT law is usually valid for the temperature dependence of the relaxation
time of the $\alpha$-process or the normal mode~\cite{Boese1990}, and
the slowing 
down of the dynamics when approaching the Vogel temperature is
essential to glassy dynamics. Therefore, we can expect that {\it
the molecular motion of the EP-process is strongly associated with the
molecular motion, the segmental motion or the normal mode, of polyamide
copolymer systems.}  
This result is consistent with the results reported in Ref.~\cite{Lu2006}.

In Fig.~\ref{fmax-T-EP-1}, there seems to be the crossing of the
temperature dependence of the rate of the EP-process $f_{\rm ep}$ and
that of the $\alpha$-process $f_{\alpha}$ especially for smaller values
of $d$. In the region where the two relaxation rates $f_{\rm ep}$ and
$f_{\alpha}$ come across, it is very difficult and almost impossible to
seperatre the component of the $\alpha$-process from that of the
EP-process. The extrapolated curve of the $\alpha$-process using the
VFT-law intersects with the observed curve of the EP-process in the
low-frequency region. Because the $\alpha$-process induces the
EP-process, it is reasonable to imagine that the two curves will merge
without intersection.

\begin{figure}
\includegraphics[width=8.3cm,angle=0]{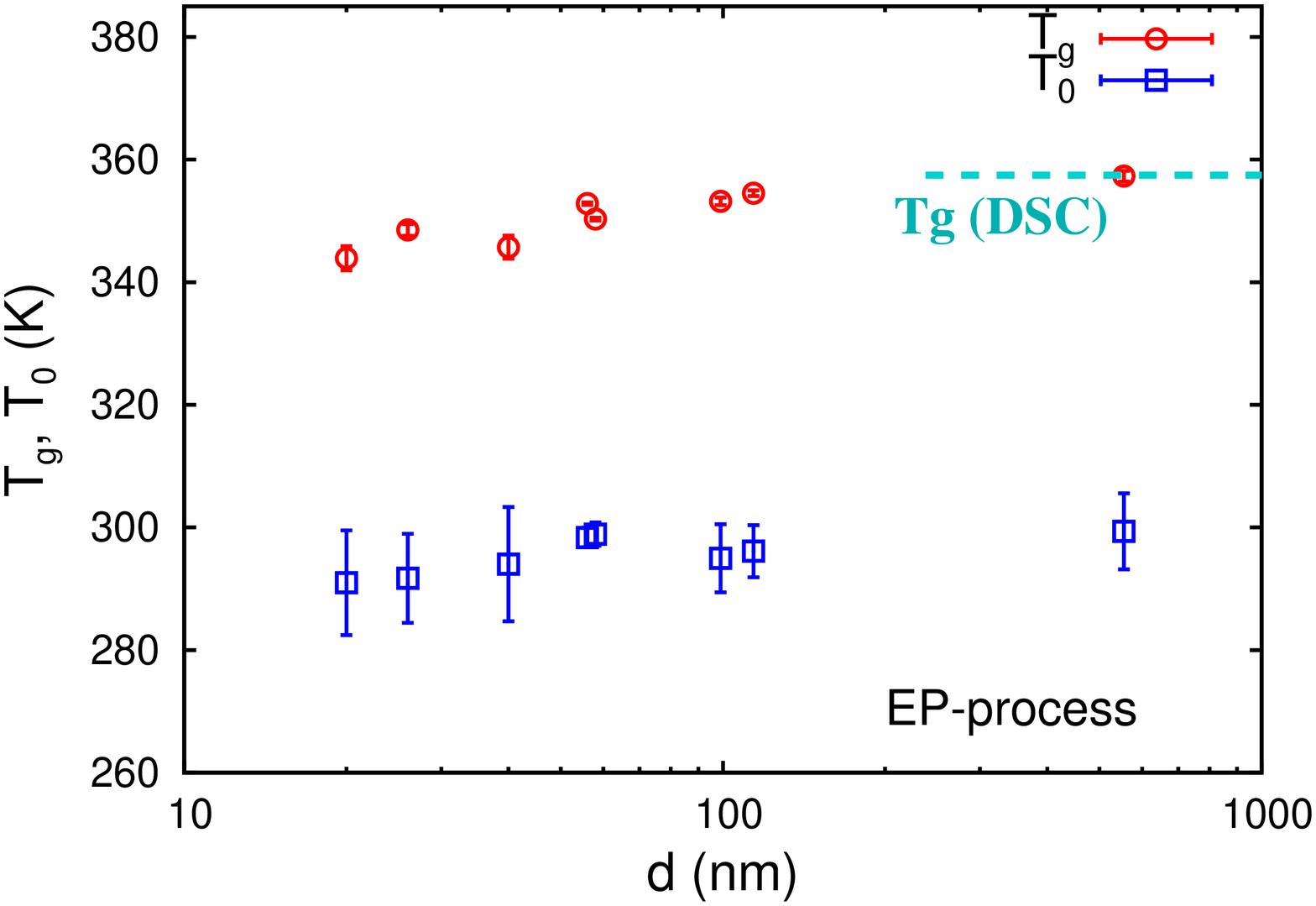}
\caption{%
The dependence of the Vogel temperature $T_0$ (\Blue{$\Box$}) and
 the glass transition temperature $T_{\rm g}$ (\Red{$\circ$}) on the
 film thickness, as determined by the temperature 
 dependence of the relaxation time $\tau_{\rm ep}$ of the EP-process for
 thin films of the amorphous polyamide copolymer 
 PA66/6I. Here, $T_{\rm g}$ is defined such that $\tau_{\rm ep}(T_{\rm
 g})=10^3$sec. The value of $T_{\rm g}$ as evaluated by DSC measurements
 is also shown. 
}\label{tg-t0-d-1}
\end{figure}

Figure \ref{tg-t0-d-1} shows the dependence of the Vogel temperature
$T_0$ on the film thickness, evaluated by fitting the observed temperature 
dependence of the relaxation rate of the EP-process to the VFT law.
Here, the glass transition temperature $T_{\rm g}$
is also evaluated, such that the relaxation time of the EP-process, 
$\tau_{\rm ep}$, is equal to 10$^3$ sec at $T_{\rm g}$, and the resulting
$T_{\rm g}$ is plotted in Fig.~\ref{tg-t0-d-1}. Furthermore, 
the glass transition temperature determined by DSC measurements is also
plotted, for comparison with other $T_{\rm g}$ results. 
Here, for the bulk system,  the $T_{\rm g}$ from DSC is 357~K, the
$T_{\rm g}$ evaluated from the relaxation rate of the EP-process 
is 357.3$\pm$0.8~K, and the Vogel temperature $T_0$ is 299$\pm$6 K.
The $T_{\rm g}$ evaluated from the EP-process agrees very well with that
from the DSC measurements. The difference between $T_0$ and $T_{\rm g}$ 
is approximately $-$58~K, which is consistent with the empirical relation that 
$T_0$ is approximately 50~K lower than $T_{\rm g}$~\cite{Strobl2007}.
Therefore, this result suggests that the relaxation rate of the
EP-process can be utilized to determine the dynamics of the
segmental  
motion, that is, the glassy dynamics, especially near the glass
transition region. 
In Fig.~\ref{tg-t0-d-1}, we can see that both $T_{\rm g}$ and $T_0$ 
decrease slightly with decreasing film thickness. 

%
%
%
%

As shown in the above, the temperature dependence of the relaxation time
of the EP-process is used to determine the glass transition
temperature $T_{\rm g}$ of the amorphous polyamide copolymer
PA66/6I. Furthermore, we can use the observed temperature dependence of
$\tau_{\rm ep}$ to determine the fragility index, which 
characterizes the glassy dynamics of amorphous materials. Here, we
define the fragility index $m$ as follows~\cite{Bohmer1992}:
\begin{eqnarray}
m&=&\left(\frac{d\log_{10}\tau_{\rm ep}(T)}{d(T_{\rm
     g}/T)}\right)_{T=T_{\rm g}},
\end{eqnarray} 
where the relaxation time $\tau_{\rm ep}$ is used instead of
$\tau_{\alpha}$, which should be used in the formal definition of the
fragility index. 

\begin{figure}
\includegraphics[width=5.9cm,angle=-90]{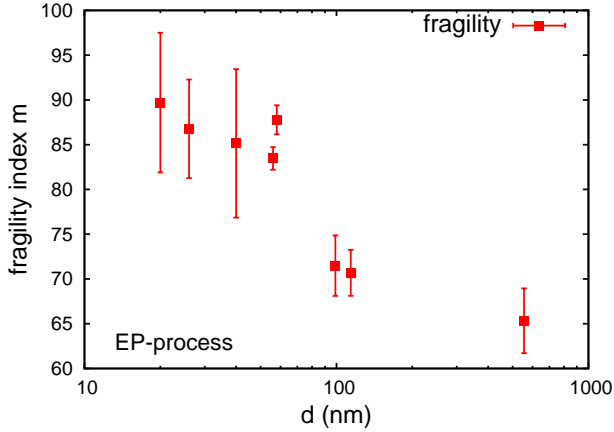}
\caption{%
The dependence of the apparent fragility index $m$ on the film
 thickness, based on the 
 temperature dependence of the relaxation time of the electrode
 polarization process. Here, $T_{\rm g}$ is defined as the temperature
 at which the relaxation time $\tau_{\rm ep}$ is equal to 10$^3$ sec.
 }\label{fragility-d-1}
\end{figure}

Figure \ref{fragility-d-1} shows the dependence of the
apparent fragility index on the film thickness, evaluated using the
temperature dependence of 
the relaxation time of the EP-process for thin films of the amorphous
polyamide copolymer PA66/6I. In Fig.~\ref{fragility-d-1}, the fragility
index increases with decreasing film thickness.
Therefore, the glassy dynamics of these thin
films are expected to become more fragile with decreasing film
thickness, provided that this 
apparent fragility index can be regarded as the fragility index 
evaluated from the relaxation time of the $\alpha$-process. A more
in-depth discussion on this issue is given in Sec.~\ref{discussions}.

\begin{figure}
\includegraphics[width=5.9cm,angle=-90]{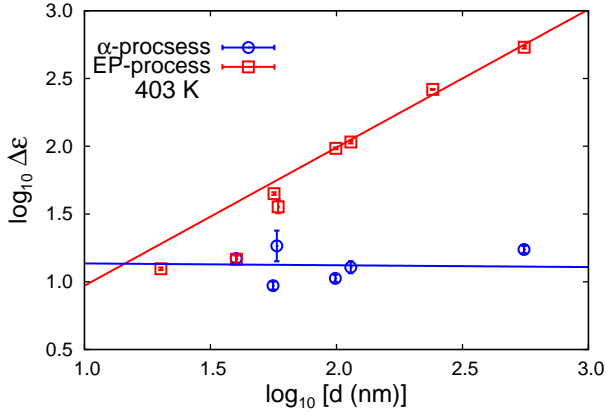}
\caption{%
The dependence of the dielectric relaxation strength
 $\Delta\varepsilon$ at 403~K on the film thickness, for the EP-process
 (\Blue{$\circ$})  and the  $\alpha$-process (\Red{$\Box$}) in thin
 films of the amorphous  polyamide copolymer 
 PA66/6I with various thicknesses. The slope of the straight line for
the EP-process is equal to unity and that for the $\alpha$-process is
 almost equal to zero.   
}\label{dielec-strength-1}
\end{figure}

\begin{figure}
\hspace*{-4.25cm}
\includegraphics[width=3.3cm,angle=-90]{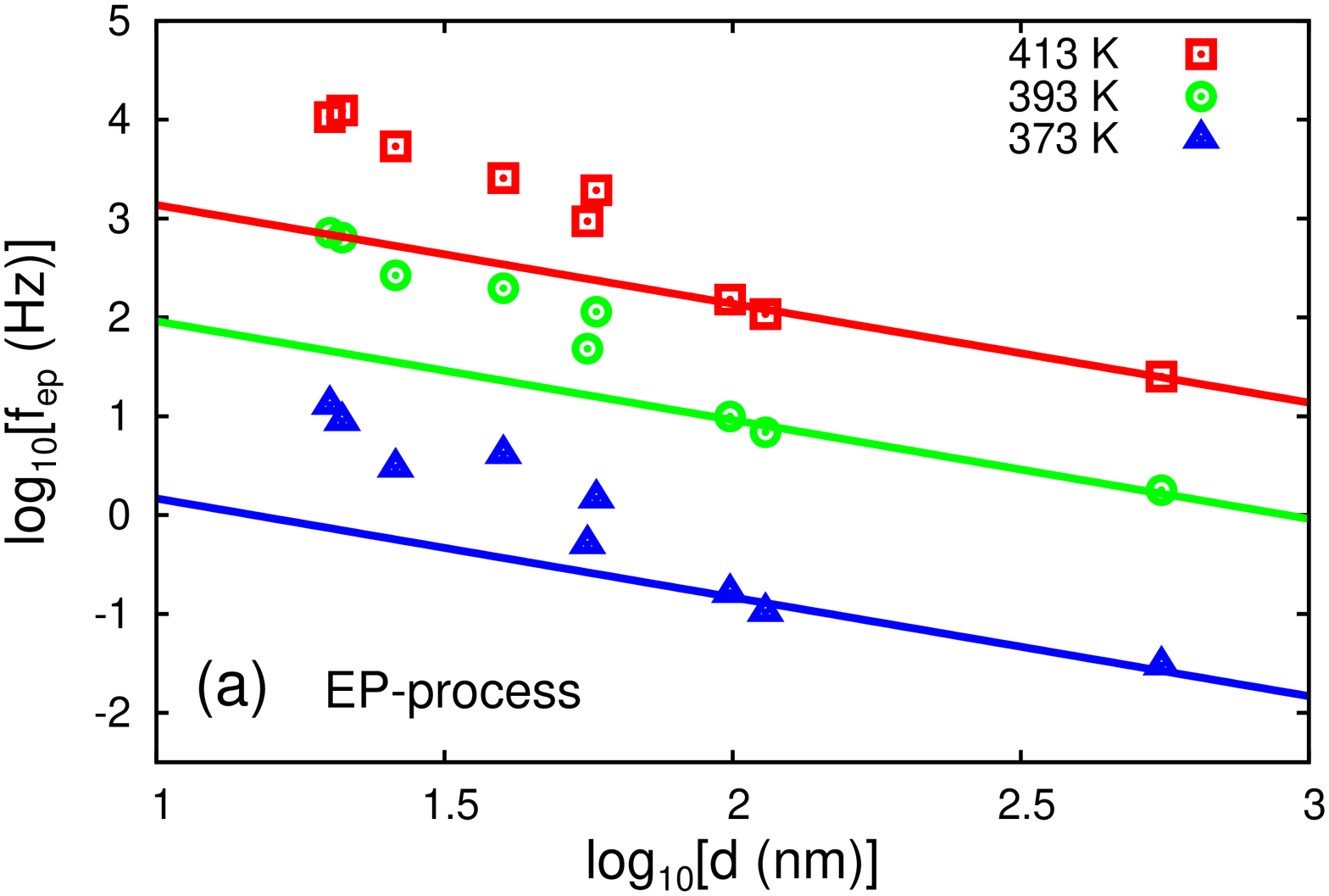}

\vspace*{-3.3cm}\hspace*{4.2cm}
\includegraphics[width=3.3cm,angle=-90]{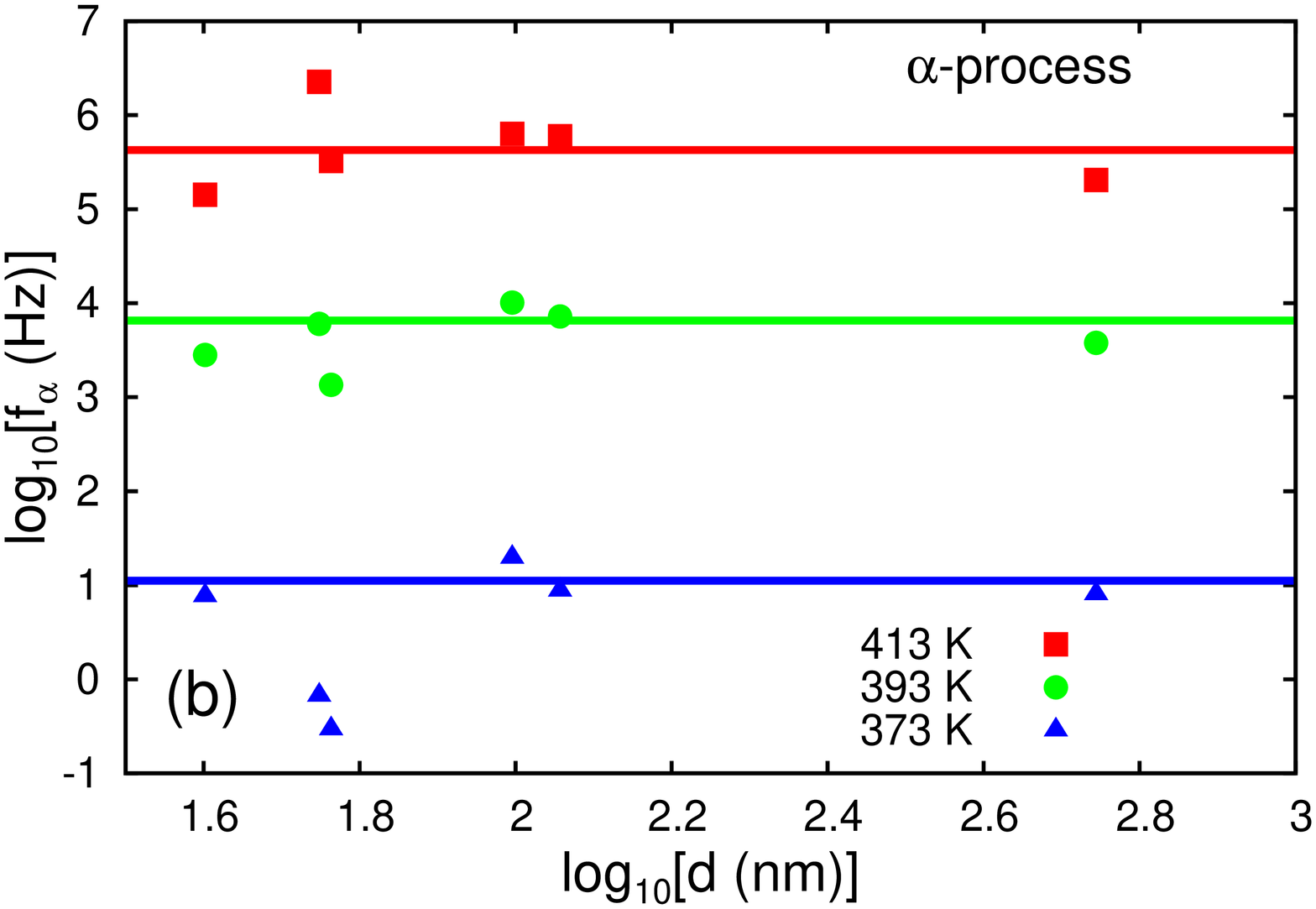}
\caption{%
 The dependence of (a) the relaxation rate $f_{\rm ep}$ of the
 EP-process and (b) the relaxation rate $f_{\alpha}$ of the
 $\alpha$-process on the film thickness, at 413~K, 393~K, and 373~K, for
 thin films of PA66/6I 
 with various thicknesses ranging from 556~nm to 20~nm. 
 The straight lines in (a) were obtained by fitting for three values of
 the thin films of PA66/6I with a thickness larger than 99~nm. The slopes 
 are fixed to be -1.
}\label{fmax-EP-alpha-d-1}
\end{figure}

The dependence of the dielectric relaxation strength on the film
thickness at 403~K for the EP-process is shown in
Fig.~\ref{dielec-strength-1}, for thin films 
of the amorphous polyamide copolymer PA66/6I. The dielectric relaxation
strength of the EP-process, $\Delta\varepsilon_{\rm ep}$, monotonically
increases with increasing film thickness. Above $d$ = 100~nm,
the thickness dependence of $\Delta\varepsilon_{\rm ep}$ is well fitted
by a straight line with a slope of unity, as shown by the red 
line in Fig.~\ref{dielec-strength-1}. According to the theoretical
model of the electrode polarization, the dielectric relaxation
strength of the EP-process can be expressed as a linear function of the
thickness of the sample, {\it i.e.}, the distance between the two electrodes, as
shown in Eq.~(\ref{del_eps_ep}).
Therefore, the linear relationship between $\Delta\varepsilon_{\rm ep}$
and the film thickness can be regarded as strong evidence for the validity
of attributing the EP-process to the strongest
dielectric loss signal observed in the present measurements.
Furthermore, it should be noted that there is a distinct deviation of
the thickness dependence of $\Delta\varepsilon_{\rm ep}$ 
from the straight line with a slope of unity. 
If $\Delta\varepsilon_{\rm ep}$ ($\tau_{\rm ep}$) can be expressed as a
linear function of $d$, the Debye length $L$ and the characteristic
relaxation time of the electrode polarization $\tau$ should be
independent of the film thickness. However, if this is not the case, the
values of $L$ and $\tau$ should depend on the film thickness. Therefore,  
this deviation implies
that {\it there is an intrinsic dependence on the film thickness for the
physical 
mechanism of the EP-process, for thin films with a thickness less than
a critical thickness $d_c$} (60~nm $<d_c<$ 100~nm ).

Figure \ref{fmax-EP-alpha-d-1}(a) shows the dependence of the
relaxation rate of the EP-process, $f_{\rm ep}$, on the film thickness,
at 413~K, 393~K, and 373~K, for thin films of the amorphous polyamide
copolymer PA66/6I. For a  
thickness larger than $d_c$, the dependence of the relaxation
rate of the EP-process on the film thickness is well described by the
following relation:
\begin{eqnarray}
f_{\rm ep}\sim d^{-1}.
\end{eqnarray}
This result is consistent with Eq.~(\ref{tau_ep})
for the present model of the EP-process. Below $d_c$,
there is an intrinsic deviation from Eq.~(\ref{tau_ep}) in the 
same manner as observed for the dielectric relaxation strength of the
EP-process.

\subsection{The $\alpha$-process}

In this section, we will discuss the experimental results on the relaxation
time of 
the $\alpha$-process for thin films of the amorphous polyamide copolymer
PA66/6I. The dielectric relaxation strength of the $\alpha$-process is
much smaller than that of the EP-process, as shown in
Fig.~\ref{eps-555nm}. It is thus difficult to evaluate the physical
parameters of the $\alpha$-process by extracting the component caused by
the $\alpha$-process from the overall dielectric spectra. However, data
fitting by Eq.~(\ref{com_eps}) enables the extraction of only the 
component from the $\alpha$-process, as well as the elucidation of the
temperature 
dependence of the relaxation rate of the $\alpha$-process, as shown in
Fig.~\ref{fmax-T-EP-1}. 

For thin films of PA66/6I with a given film thickness, the relaxation rate
of the $\alpha$-process is located at a higher frequency region than
that of the EP-process, at a given temperature. The deviation of the
relaxation rate of the $\alpha$-process from that of the EP-process    
becomes smaller as the temperature approaches the glass transition
temperature. Hence, the $\alpha$-process appears to merge with the
EP-process with decreasing temperature. This is the reason for the
difficulty in evaluating the relaxation rate of the $\alpha$-process.
At the same time, the merging of the $\alpha$-process with the EP-process
can be regarded as an evidence that the EP-process is strongly
associated with the $\alpha$-process, especially at the glass transition
region. 
In Fig.~\ref{fmax-T-EP-1}, the temperature dependence of the relaxation
rate of the $\alpha$-process is well reproduced by the VFT law, 
under the condition that the Vogel temperature of the $\alpha$-process
is the same as that of the EP-process for each film thickness.  

As for the dependence of the $\alpha$-process on the film thickness, the
dielectric relaxation strength and the relaxation rate of the
$\alpha$-process are shown as functions of the film thickness at a given
temperature in Fig.~\ref{dielec-strength-1} and
Fig.~\ref{fmax-EP-alpha-d-1}(b), respectively.  
Although there is some scatter in the data points, they suggest that
there is almost no systematic dependence of the 
dielectric relaxation strength and the relaxation rate of the
$\alpha$-process on the film thickness, for the thickness range
investigated in this study. 
The dependence of the relaxation rate of the $\alpha$-process on the
film thickness in comparison to that of the EP-process is discussed in
Sec.~\ref{discussions}, for thin films of PA66/6I.

\begin{figure}
\includegraphics[width=8.4cm,angle=0]{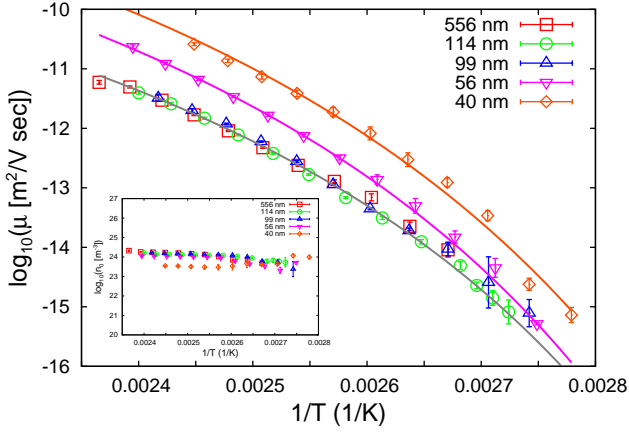}
\caption{%
The dependence of the charge carrier mobility $\mu$ on the inverse
 of the absolute temperature $1/T$, for thin films of the amorphous polyamide
 copolymer PA66/6I with various thicknesses ranging from 556~nm to 40~nm.  
The curves were obtained by fitting the observed values to the VFT
 equation for the mobility, Eq.~(\ref{mu}). In the inset, the dependence of
 the equilibrium concentration of charge carriers $n_0$ on $1/T$ 
 is also given for the same thin films of the amorphous polyamide
 copolymer PA66/6I.
 }\label{mu_n0}
\end{figure}

\subsection{Relaxation of the distribution of space charge}

As shown in Sec.~\ref{ep-process}, the dielectric relaxation phenomenon
of the EP-process exhibits an interesting dependence on the temperature
and film thickness, for the amorphous polyamide copolymer PA66/6I.
Here, we try to evaluate the intrinsic nature of the EP-process, based
on the theoretical model given 
in Sec.~\ref{ep-process-theory}.
From the observed values of $\tau_{\rm ep}(T)$ and
$\Delta\varepsilon_{\rm ep}$,  
the temperature dependence of the mobility and equilibrium concentration
of charge carriers within  
the polymeric system can be evaluated using Eqs.~(\ref{mu}) and (\ref{n0}).
Figure \ref{mu_n0} shows this dependence for thin films of PA66/6I of
various thicknesses. In Fig.~\ref{mu_n0}, the equilibrium 
concentration of charge carriers $n_0$ shows a weak temperature
dependence (figure inset), while the mobility $\mu(T)$ 
shows a strong temperature dependence that can be expressed by the VFT
law of mobility as follows:
\begin{eqnarray}\label{VFT_mu}
\mu &=& \mu_0\exp\left(-\frac{\tilde{U}}{T-T_0}\right).
\end{eqnarray}
Here, $\mu_0$ and $\tilde{U}$ are positive constants. The Vogel
temperatures $T_0$ obtained from $\mu(T)$ are 294~K and 298~K for
$d=$40~nm and $d>$99~nm, respectively. These values are nearly identical
to those evaluated from $\tau_{\rm ep}(T)$, as shown in 
Fig.~\ref{tg-t0-d-1}.

For the thin films with thicknesses of 556~nm, 114~nm,
and 99~nm, the temperature dependences of the mobility agree well with
each other, and are well reproduced by the VFT law with the
same parameters. However, for the thin films with thicknesses less than
$d_c$, the mobility of the charge carriers at a given temperature
increases with decreasing film thickness. This is consistent with the results
of the relaxation rate of the EP-process in
Fig.~\ref{fmax-EP-alpha-d-1}, that is, the crossover of $f_{\rm ep}$ 
from the straight line with a slope of -1 to the line with a larger
slope as the film thickness decreases. Therefore, we can conclude that
{\it the intrinsic mobility of the charge carriers remains almost unchanged 
for a film thickness above a critical thickness $d_c$},  
{\it while the mobility increases with
decreasing film thickness below $d_c$.}

\begin{figure}
\includegraphics[width=6.0cm,angle=-90]{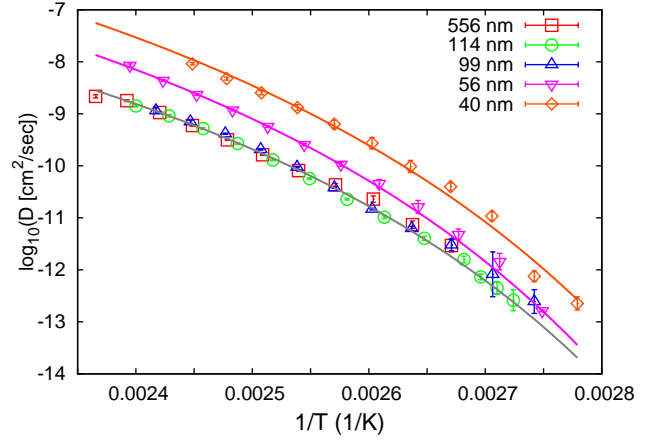}
\caption{The dependence of the diffusion constant of the charge carriers $D$ on
 the inverse of the absolute temperature $1/T$, for thin films of the
 amorphous polyamide copolymer PA66/6I with various thicknesses ranging
 from 556~nm to 40~nm.   
}\label{diffusion_const}
\end{figure}

Furthermore, the diffusion constant of the charge carriers $D(T)$ can be
evaluated from the mobility $\mu$(T) using Eq.~(\ref{diffusion_1}), as
shown in Fig.~\ref{diffusion_const}. In Fig.~\ref{diffusion_const}, the
diffusion constant can be reproduced using the relation
\begin{eqnarray}\label{vft_diffusion}
D(T)&=&D_0T\exp\left(-\frac{\tilde{U}}{T-T_0}\right), 
\end{eqnarray}
where $D_0$ is a positive constant. The value of $D$ changes from
10$^{-13}$ cm$^2$/sec to 10$^{-8}$cm$^2$/sec as the temperature changes
from 363~K to 410~K, for thin films of the amorphous polyamide copolymer
PA66/6I with a thickness larger than $d_c$. Below $d_c$, an intrinsic
increase in $D$ is observed, similar to that for the mobility $\mu$.
Therefore, the translational motion of the charge carriers is
intrinsically enhanced with decreasing film thickness.

\section{Discussion}
\label{discussions}

From the
observed dielectric spectra for thin films of PA66/6I, we successfully
discerned the separate contributions from the EP-process and the
$\alpha$-process.   
The physical origin of the $\alpha$-process is due to segmental 
motion, which can be regarded as a rotational motion of polymeric
segments, while that of the EP-process is due to the translational motion
of charge carriers. Because the motion of charge carriers is activated
via the molecular motion of the polyamide systems in which the charge
carriers are located, the motion of charge carriers is strongly
associated with the molecular motions of polymer chains.  

\subsection{Glassy dynamics evaluated from the EP-process}

As shown in Fig.~\ref{fmax-T-EP-1}, the temperature dependence of
$\tau_{\rm ep}(T)$ is well reproduced by the VFT law. Hence, we expect
that the glass transition behavior of PA66/6I can be 
evaluated from the EP-process. Figure \ref{tg-t0-d-1} clearly shows that
$T_{\rm g}$ is well evaluated using $\tau_{\rm ep}(T)$. The
fragility index evaluated from $\tau_{\rm ep}$ increases with decreasing
film thickness, as shown in Fig.~\ref{fragility-d-1}. The 
dependence of $T_{\rm g}$ and the fragility index $m$ on the film
thickness has been measured
for the thin films and/or nanoparticles of several polymers without strong
polarity~\cite{Fukao2001a,Evans2013,Zhang2013}. Both $T_{\rm g}$ and
the fragility index decrease with decreasing film thickness for many
cases. In contrast, the present result clearly shows that as the
film thickness decreases, the fragility index increases, while 
$T_{\rm g}$ decreases. This inverse dependence on thickness of $m$ might
be associated with the strong polarity of PA66/6I. The fragility
index can be regarded as a measure of the cooperativity of the
$\alpha$-process or segmental motion~\cite{Bohmer1992}. In the case of
polyamides, the formation of a network as a result of the hydrogen
bonding can promote the cooperativity of the glassy dynamics, even in
thin films.    

\subsection{Coupling or decoupling of translational and rotational
  motions} 

As shown in Fig.~\ref{fmax-EP-alpha-d-1}(a), the relaxation rate of the
EP-process shows an intrinsic increase below a critical thickness, that
is, $f_{\rm ep}$ increases with decreasing film thickness with a
stronger thickness dependence than that described by the linear
relationship between $f_{\rm ep}$ and $d^{-1}$. 
This corresponds to the enhancement in the diffusion motion of the
charge carriers in the thin films with $d<d_c$, as shown in 
Fig.~\ref{diffusion_const}. On the other hand, the relaxation rate of
the $\alpha$-process shows almost no systematic dependence on the film
thickness, for the thickness range investigated in this study, although
there is scatter in the data points. This might be experimental evidence
for the decoupling of the rotational motion of polymer chain segments
and the translational motion of charge carriers in amorphous polyamide    
copolymers. More precise measurements are required in
order to discuss this decoupling of the rotational motion and
the translational motion in greater detail.

\subsection{Origin of charge carriers}

The dielectric relaxation spectra of amorphous polyamide copolymers
clearly show strong signals for the EP-process. 
As mentioned in Sec.~\ref{introduction}, there are several possible
candidates for 
charge carriers in polyamides. One of the most promising is ionized
protons. From the present analysis of the EP-process on the 
basis of the theoretical model developed by Coelho~\cite{Coelho1983}, the
diffusion constant and equilibrium concentration of the charge carriers
are obtained as shown in Figs.~\ref{mu_n0} and
\ref{diffusion_const}. The charge carrier concentration $n_0(T)$ shows
almost no significant temperature dependence for the temperature range
investigated in this study, and has a value of $n_0\approx 10^{18}$
cm$^{-3}$. The concentration of hydrogen atoms in the polyamide can be
evaluated as $n_0^H\approx 5\times 10^{21}$cm$^{-3}$. Therefore, the ratio of
the concentration of the charge carriers to that of the hydrogen atoms in
the polyamide can be calculated as follows:
$n_0/n_0^H$= 0.02~\%. Hence, ionized protons are possible
candidates for the charge carrier in PA66/6I.

The absolute values of the diffusion constants of the charge carriers
obtained in this study can be compared with those of various particles
in the literature in the following mammer:  
\begin{enumerate}
\item
It is well-known that there is a possibility of water
uptake in polyamide. The diffusion constant of adsorbed water molecules
in amorphous polyamide copolymers has been measured by neutron
scattering and is estimated to be between 10$^{-6}$ and
10$^{-5}$cm$^2$/sec for the temperature range from 350~K to
430~K~\cite{Laurati2012}.  
\item
The mutual diffusion constant of polystyrene (PS) chains at the interface
between h-PS and d-PS has been reported in Ref.~\cite{Whitlow1991}. 
For $M_{\rm w}=$10$^5$, the mutual diffusion constant of PS changes from
2.0$\times$10$^{-16}$ to 1.2$\times$10$^{-14}$cm$^2$/sec, as the
temperature increases from 396~K to 413~K.
\item
Similar measurements of the diffusion constant of deuterated polyethylene
(PE) in a matrix of hydrogenated PE show that the diffusion cosntant 
at 449~K is equal to 0.2$\times M_{\rm w}^{-2.0}$~\cite{Klein1978}.  
Here, for $M_{\rm w}=10^4$, the value of $D$ is equal to 
2$\times$10$^{-9}$cm$^2$/sec.  
\item
The tracer diffusion constant of the chloride ion Cl$^{-}$ in polyamide-6 film
has been measured in Ref.~\cite{Chantrey1969}. The diffusion constant
ranges from 1.4$\times$10$^{-9}$ to 3.4$\times$10$^{-8}$cm$^2$/sec,
depending on the pH at room temperature.
\item
The calculated diffusion constant of the hydrogen combined to the
     terminal amines in polyamide-6,6
is reported to be 6$\times$10$^{-13}$cm$^2$/sec at 298~K~\cite{Zaikov1988}.
\end{enumerate}
The diffusion motion of water adsorbed on PA66/6I is much faster than the
observed diffusion motion in this study. The observed diffusion constants of
the chloride in polyamide-6 and the calculated ones of the combined hydrogens in
polyamide-6,6 are larger than those observed in PA66/6I. 
On the other hand, the diffusion constants related to the reptation
motion of PE chains are comparable to the observed values.

The comparison of the observed diffusion constants for PA66/6I in
Fig.~\ref{diffusion_const} with various diffusion constants in the
literature suggests that the molecular motion of the charge carriers 
is strongly correlated with the motion of entire chains of
the polyamide, if ionized protons are the actual charge carriers responsible
for the EP-process in PA66/6I. The observed results are quite 
consistent with this idea, although further direct measurements of
both the molecular motion of ionized protons and the reptation motion of
the polyamide are required for further investigation.

\section{Concluding remarks}\label{conc_remarks}

In this study, the dynamics of the electrode polarization process
(EP-process) and the $\alpha$-process in thin films of an amorphous
polyamide copolymer were investigated using dielectric
relaxation spectroscopy measurements. The obtained results are
summarized as follows: 
\begin{enumerate}
\item
The relaxation time of the EP-process has a  Vogel-Fulcher-Tammann type
     of temperature dependence, and the $T_{\rm g}$ evaluated from the
     EP-process agrees very well with the $T_{\rm g}$ determined from
     differential scanning calorimetry measurements. The fragility index
     derived from the EP-process increases with decreasing film thickness.  
\item
There is a distinct deviation from this linear law for thicknesses
     smaller than a critical value. This deviation corresponds to an
     increase in the diffusion constant of the charge carriers, which are
     responsible for the EP-process. The $\alpha$-process is located in
     a region of higher frequency than the EP-process at high
     temperatures, but merges with the EP-process near $T_{\rm g}$. 
\item
The dependence of the relaxation time of the $\alpha$-process on the
     film thickness is 
     different from that of the EP-process. This suggests that
     there is decoupling between the segmental motion of the polymers
     and the translational motion of charge carriers in confinement. 
\end{enumerate} 

\vspace*{0.1cm}
In the present study, we successfully derived the physical properties of
the glassy dynamics from those of the EP-process. This means that it is
possible to evaluate the glassy dynamics even if the $\alpha$-process
cannot be observed as a result of the existence of a large signal
related to the motion of charges. Molecular motion, such as the
$\alpha$-process, may be clarified in large polar polymeric sytems
through the EP-process. For crystalline polyamide systems, the
Maxwell-Wagner-Sillar interfacial polarization process should appear, in 
addition to the EP-process. Even in such systems, the evaluation of the
glassy dynamics from the EP-process should be possible in principle,
although actual data analysis would be very difficult. 
We also showed that there is possible decoupling between the rotational
motion of polyamide chain segments and the translational motion of
charge carriers. 

\acknowledgements

This work was supported by a Grant-in-Aid for Scientific
Research (B) (No. 25287108) and Exploratory Research
(No. 25610127) from the Japan Society for the Promotion of Science. The
authors would like to express their cordial thanks to Solvay for
supplying them with the amorphous random copolymer PA66/6I. 
They also thank the anonymous reviewer for the very useful comments and
suggestions. 
The synchrotron radiation experiments were performed at the
BL40B2 of SPring-8 with the approval of the Japan Synchrotron Radiation
Research Institute (JASRI) (Proposals 2013A1173 and 2014A1230).


\end{document}